# Phase transition of the Si(111)-4×1-In surface reconstruction investigated by electron transport measurements

Takashi Uchihashi[1*] and Urs Ramsperger[2]

[1] *National Institute for Materials Science, 1-2-1, Segen, Tsukuba, Ibaraki, Japan*
[2] *Laboratorium für Festkörperphysik, Eidgenössische Technische Hochschule (ETH) Zürich, CH-8093 Zürich, Switzerland*

**Abstract**

We measure the electron conductivity of the surface states and the subsurface space charge layer originating from the Si(111)-4×1-In reconstruction as a function of temperature. The conductivity of the *surface states* drops sharply around 130 K with decreasing temperature, revealing a metal-insulator phase transition of the surface reconstruction. In contrast, the influence of the phase transition on the conductivity of the *space charge layer* is limited to temperatures above 60 K. This means that the surface Fermi level remains strongly pinned despite the phase transition, indicating the presence of free carriers in the surface states down to rather low temperatures.

*Corresponding author
E-mail address*: UCHIHASHI.Takashi@nims.go.jp

# 1. Introduction

Although surface science has developed a variety of sophisticated experimental techniques in the course of its history, electron transport properties of well-defined surface reconstructions have rarely been investigated to date. This is primarily the result of experimental techniques in surface science being sensitive to the properties of the first few layers on a surface, while macroscopic electron transport properties were generally thought to be dominated by conduction through bulk. Recently, however, surface transport phenomena have proven experimentally accessible via direct dc transport measurements,[1,2] providing a new approach to the possibility of investigating various surface systems. For example, the temperature-induced phase transition of the Si(111)-4×1-In reconstruction observed around 130 K [3] should benefit from electron transport studies. This system exhibits a metal-insulator transition with lattice and charge modulations, which is attributed to the Peierls instability originating from the 1D nature of the system.[4] Still, several subsequent studies[5-7] do not fully support this simple picture of the Peierls transition. We may conclude, however, that detailed analysis of conductivity measured as a function of temperature could clarify the nature of this phase transition.

We have recently reported the electron conduction of the Si(111)-4×1-In surface reconstruction over a wide temperature range.[8] The metal-insulator transition was confirmed through measurements of the surface state conductivity for the first time, and this paper provides analysis of the temperature dependence of conductivity of both the surface states and the space charge layer. The conductivity of the *surface states* drops sharply around 130 K with decreasing temperature, followed by an asymptotic approach toward zero. In contrast, the phase transition does not qualitatively change the conduction through *the space charge layer* when the temperature is above 60 K. Despite the phase transition, the Fermi level of the surface states is strongly pinned. This indicates the presence of free carriers in the surface states and, consequently, the role of large thermal fluctuation unique to the 1D electron system.

# 2. Experiment

All experiments are performed under ultrahigh vacuum (UHV) conditions.

Non-doped Si(111) (resistivity ρ> 1000 Ωcm) is chosen as the substrate to minimize the electron conduction through the bulk. Two tantalum electrode pads separated by 1 mm are deposited beforehand on the sample with an electron beam evaporator. After loading into the UHV chamber, the samples are cleaned by flashing at 1150 °C for 30 s. Indium is then deposited and the sample is annealed around 450 °C for 5 minutes. This induces the Si(111)-4×1-In reconstruction (Fig. 1(a)).[3,5,6] The 4×1 reconstruction is confirmed over an extensive area by Low Energy Electron Diffraction (LEED), while local structures such as point defects and surface steps are observed with a scanning tunneling microscope (STM) at room temperature.

*In situ* dc two-probe measurements are conducted after STM observation. Two spring-loaded gold-coated probes are pressed onto the electrodes to ensure stable and reliable electrical contacts (Fig.1(b)). The conductivity is measured between room temperature and 6 K. The carriers in the bare silicon substrates are found to be quenched below 220 K. This allows the selective detection of electrical conduction originating from the surface structures: the possible conduction path is either through the surface states or the underlying subsurface space charge layer. The current-voltage characteristics are linear for -1 V < $V$ < 1 V over a wide temperature range, confirming the absence of a Schottky barrier at the electrode interfaces.

## 3. Results and discussions

Figure 2(a) shows the temperature dependence of conductivity of a sample with the Si(111)-4×1-In reconstruction in two consecutive measurements. First, a well-developed Si(111)-4×1-In reconstruction is prepared with negligible native defects. The conductivity ($\sigma$) shows a significant decrease around 130 K with decreasing temperature (solid line), which is attributed to decrease in surface conductivity caused by the metal-insulator transition. After returning to room temperature, a small amount of indium (less than 0.1 ML) is additionally deposited without any sample annealing. This process introduces defect-like structures in the middle of the indium chains (Fig. 2(b)). These "defects" are approximately 0.5 Å in depth in STM images, clearly distinct from the intrinsic point defects with a depth of about 1 Å. In the second conductivity measurement (Fig. 2(a), dotted line), $\sigma$ significantly decreases compared to the first above 130 K, while the two are almost identical in the measurements below 100 K. Because electron conduction is suppressed by defects in the indium chains, the difference between the two (shaded area) is the contribution of the surface reconstruction. The conductivity of the surface states ($\sigma_{surf}$) obtained this way is plotted with a dotted-dashed line in Fig. 2(a) (five-times magnified). $\sigma_{surf}$ exhibits a sharp decrease at 125 K (defined as $T_c$) with decreasing temperature, followed by an asymptotic approach toward zero. We find that $\sigma_{surf} > 0$ at least down to 80 K.

We analyze the data by plotting the quantity $-d(\ln\sigma)/d(1/k_B T)$ as a function of $T$, where $k_B$ is the Boltzmann constant. Figure 3(a) shows the $-d(\ln\sigma)/d(1/k_B T)$-$T$ plot for the first (solid line) and second (dotted line) measurements. A sharp peak observed just below 125 K in the first measurement disappears in the second, while the two plots are nearly identical in the other region. This means that surface conduction is completely suppressed by introduction of defects, because the peak is the signature of the metal-insulator transition of the surface reconstruction. Hence, the conductivity measured in the second measurement is attributed only to the space charge layer (denoted as $\sigma_{sc}$ in the following).

The $-d(\ln\sigma_{sc})/d(1/k_B T)$-$T$ plot is nearly linear between 200 K and 60 K. This is equivalent to the fact that $\sigma_{sc}$ is represented by the following equation

$$\sigma_{sc} = C \exp(-E_{f\text{-surf}}/k_B T) T^{-\alpha}, \quad (1)$$

where $C$ is a constant. $E_{f\text{-surf}}$ is the surface Fermi level measured relative to the upper bound of the valence band at subsurface (and hence an activation energy for holes in the subsurface space charge layers; see the inset of Fig. 3(a)), and the term $T^{-\alpha}$ corresponds to the power-law $T$-dependence of the carrier mobility.[9] Since the $-d(\ln\sigma_{sc})/d(1/k_B T)$-$T$ plot is slightly bent at 125 K, we analyze the data for the high temperature (HT) regime (130 K < $T$ < 170 K) and the low temperature (LT) regime (70 K < $T$ < 120 K), separately. By curve fitting to Eq. (1), $E_{f\text{-surf}}$ = 45 meV, $\alpha$ = 3.5 for the HT regime and $E_{f\text{-surf}}$ = 35 meV, $\alpha$ = 2.6 for the LT regime. Figure 3(b) shows the temperature dependence of conductivity calculated using parameters obtained for the HT regime (dotted line), along with the measurement (solid line). We see that the phase transition slightly shifts $\sigma_{sc}$ in the positive direction below $T_c$. (footnotes)

Although the surface Fermi level is affected by the phase transition, it remains much lower than the corresponding value deep within the bulk substrate (~ 0.5 eV). In other words, the Fermi level of the surface states remains strongly pinned as temperature decreases across $T_c$ down to about 60 K. This indicates the presence of free carriers in the surface states in this temperature range.[10] This may seem inconsistent with the metal-insulator transition of the surface reconstruction, but it is reasonable considering the large thermal fluctuation in the 1D electron system.[4,6] In this case, free carriers survive until enough low temperatures compared to $T_c$. Indeed, the steep rise in $-d(\ln\sigma_{sc})/d(1/k_BT)$ (equivalently, rapid decrease in $\sigma_{sc}$) below 60 K could be attributed to disappearance of the surface Fermi level pinning. This is reasonable considering the temperature dependence of the surface conductivity described above; free carriers in the surface states decrease sharply around 130 K, but remain observable until 80 K. It is expected to extinguish only below 60 K. Thus, detailed analysis of electron conduction through the *space charge layer* could give important information on the carrier density of the *surface states* below $T_c$.

## 4. Summary

In summary, we have investigated the phase transition of the Si(111)-4×1-In surface reconstruction by observing electron transport phenomena. The conductivity of the surface states and that of the space charge layer were separated by intentionally introducing defects in the surface indium chains. The conductivity of the surface states drops sharply around 130 K with decreasing temperature, revealing a metal-insulator phase transition of this surface reconstruction. In contrast, the phase transition does not qualitatively change the conduction through *the space charge layer* for $T > 60$ K. This shows that the surface Fermi level remains strongly pinned despite the phase transition. The large thermal fluctuation unique to the 1D electron system may explain the presence of free carriers in the surface states down to rather low temperatures. Detailed analysis of electron conduction through the space charge layer could provide critical insight into the nature of the phase transition in the surface states.

This work was supported by the Science and Technology Agency through Organized Research Combination System.

# Footnotes

In Ref. 8, an excellent fitting of $\sigma$ to Eq. (1) was obtained over a wide temperature range (Fig. 1(d) in Ref. 8) for a sample with defects introduced *at step edges*. In this sample, suppression of surface conduction due to the defects at steps is imperfect; the -d(ln$\sigma$)/d(1/$k_B T$)-$T$ plot shows a minor peak at 130 K, indicating the presence of small surface conduction. The two effects, *i.e.*, positive deviation of $\sigma_{sc}$ from the high temperature behavior and decrease in $\sigma_{surf}$ below $T_c$, cancel out each other, giving an apparent perfect fitting.

**Figure Captions**

Fig.1 (a) Representative STM image of the Si(111)-4×1-In surface reconstruction. The sample bias voltage is –1.5 V (filled state image). (b) Diagram of the electron conduction measurements of the Si(111)-4×1-In reconstruction. After sample characterization, *in situ* conduction measurements are taken under UHV environment. Two spring-loaded gold-coated probes are pressed onto the Ta electrodes, assuring stable and reliable electrical contacts.

Fig. 2 (a) Temperature dependence of conductivity of a sample with the Si(111)-4×1-In reconstruction in two consecutive measurements. Solid line: first measurement. Dashed line: second measurement after introduction of defects in the middle of the indium chains. The difference between the two measurements is the contribution from the surface reconstruction (dotted-dashed line, five-times magnified). (b) STM image of the sample used for the second measurement. Indium-induced defects are introduced in the middle of the chains. The sample bias voltage is –1.5 V (filled state image).

Fig. 3 (a) -d(ln$\sigma$)/d(1/$k_B T$)-$T$ plot for the first (solid line) and second (dotted line) measurements. A sharp peak observed at $T_c$ in the first measurement disappears in the second, while two plots are nearly identical in the other region. The temperature regions used for curve fitting are also shown. Inset: the energy band diagram of the system of the surface Si(111)-4×1-In reconstruction and the silicon substrate. $E_{\text{f-surf}}$ is defined as the surface Fermi level measured relative to the upper bound of the valence bad at subsurface. (b) Temperature dependence of the space charge layer conductivity. Solid line: measurement. Dotted line: calculation based on fitting parameters obtained for the HT regime.

Fig.1

(a)

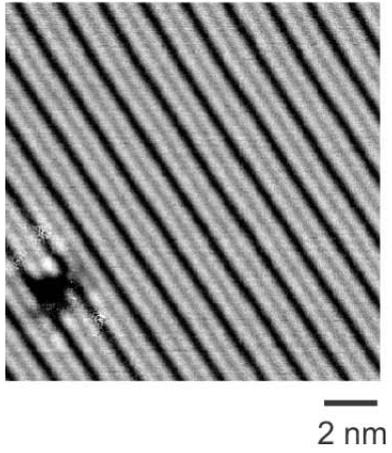

2 nm

(b)

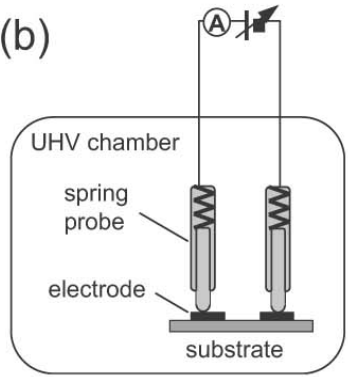

Fig. 2

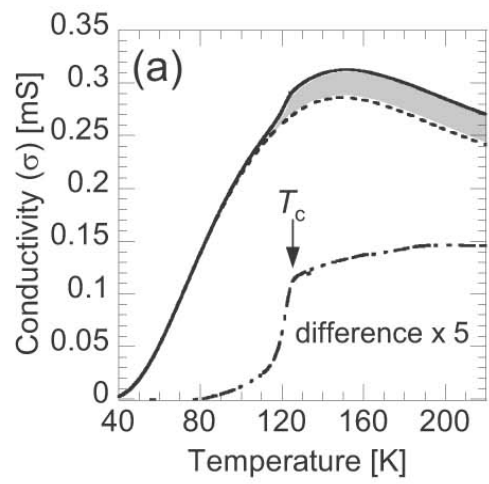

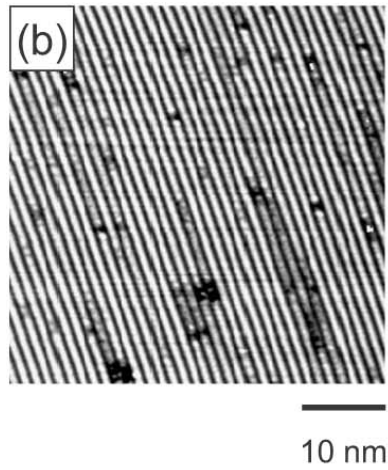

10 nm

Fig. 3

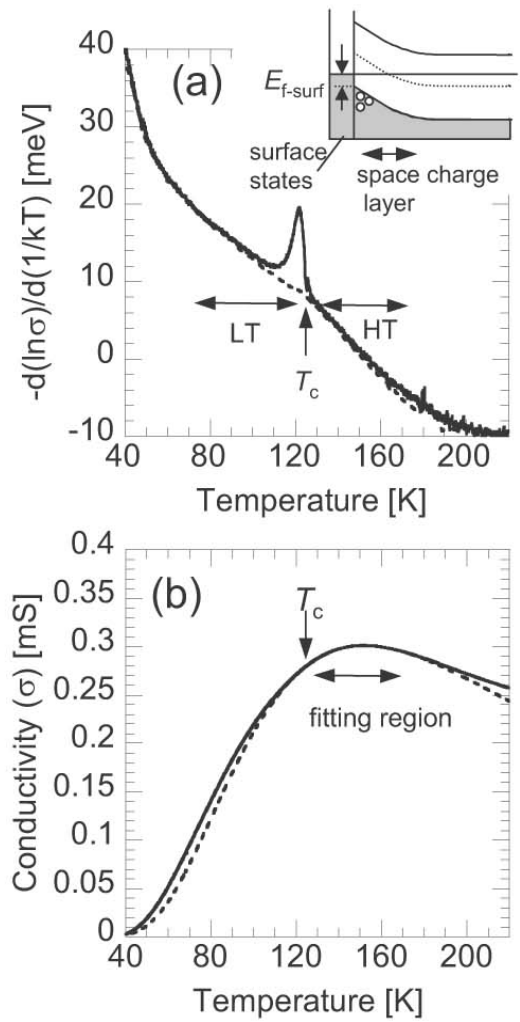